\documentclass[twocolumn]{revtex4}
\usepackage{graphicx}
\usepackage{dcolumn}
\usepackage{longtable}
\usepackage{color}

\begin {document}

\title{On the lifetime of the $2_1^+$ state in $^{10}$C}

\author{E.A. McCutchan,$^{1,2}$ C.J. Lister,$^1$ Steven C. Pieper,$^1$
  R.B. Wiringa,$^1$ D. Seweryniak,$^1$ J. P. Greene,$^1$
  P.F. Bertone,$^1$ M.P. Carpenter,$^1$ C.J. Chiara,$^{1,3}$
  G. G\"{u}rdal,$^4$ C.R. Hoffman,$^1$ R.V.F. Janssens,$^1$
  T.L. Khoo,$^1$ T. Lauritsen,$^1$ and S. Zhu$^1$}

\affiliation{$^1$ Physics Division, Argonne National Laboratory, Argonne, Illinois 60439, USA}

\affiliation{$^2$ NNDC, Brookhaven National Laboratory, Upton, New York 11973, USA}

\affiliation{$^3$Department of Chemistry and Biochemistry, University of Maryland, College Park, Maryland 20742, USA}

\affiliation{$^4$Nuclear Engineering Division, Argonne National Laboratory, Argonne, Illinois 60439, USA}

\begin {abstract}


The lifetime of
the J$^{\pi}$=$2_1^+$ state in $^{10}$C was measured using the Doppler Shift Attenuation
Method following the inverse kinematics $p$($^{10}$B,$n$)$^{10}$C
reaction at 95 MeV.  The $2^+_1$ state, at 3354 keV, has
$\tau$ = 219${\pm}$(7)$_{stat}$ ${\pm}$(10)$_{sys}$ fs corresponding to a
$B(E2)\downarrow$ of 8.8(3) $e^2$fm$^4$.  This measurement, combined with
that recently determined for $^{10}$Be (9.2(3) $e^2$fm$^4$), provides a
unique challenge to {\it ab-initio} calculations, testing the structure
of these states, including the isospin symmetry of the wave functions.
Quantum Monte Carlo calculations using realistic two- and three-nucleon
Hamiltonians that reproduce the $^{10}$Be $B(E2)$ value generally predict a larger
$^{10}$C $B(E2)$ probability but with considerable sensitivity to the admixture of
different spatial symmetry components in the wave functions, and to the
three-nucleon potential used.

\end {abstract}

\maketitle

\section{Introduction}

A new generation of {\it ab-initio} calculations based on realistic nucleon-nucleon forces
has deepened our understanding of how nuclei work. Comparison of their predictions to precise
new experimental data~\cite{wuosmaa,mueller,grinyer} has guided improvements in the calculations, both in the computational methods
used and in the underlying Hamiltonians.  Electromagnetic decay rates have already proven to be surprisingly
sensitive for probing 3-body forces~\cite{mccutchan}. In this paper we examine mirror symmetry in nuclei, comparing
$^{10}$Be to $^{10}$C, the lightest $T=1$ mirror pair for which bound excited states exist. We have precisely
measured the $B(E$2;$2_1^+ \rightarrow 0_1^+$) from the only bound state in $^{10}$C and compared it to its
analog in $^{10}$Be to probe the subtleties of the symmetry. The experimental result is difficult to reproduce
using our Green's Function Monte Carlo (GFMC) method. A variety of calculations have been performed in order to
better understand the issues.

Conceptually, the $A$ = 10 mirror nuclei are interesting. $^{10}_{\:4}$Be$_6$
can be thought of as two alpha particles with isospin $T=0$, bound to a
correlated pair of $s$-wave neutrons, which are
usually outside the alpha clusters. Similarly $^{10}_{\:6}$C$_4$ consists of two alpha
particles surrounded by a loosely bound pair of protons.  Naively
one might expect the two protons to result in a bigger $B(E2)$ value for
$^{10}$C than for $^{10}$Be.  For example, a
simple classical isoscalar liquid drop model~\cite{BM} is driven by charge:
the carbon isotope has a larger quadrupole moment ($\sim$~$Z$)
and the decay strength is expected to be larger in $^{10}$C compared to
$^{10}$Be by a factor (6/4)$^2$ = 2.25. In contrast, simple shell
models~\cite{cohenandkurath,millener} always have properly constructed
quantum-mechanical wave functions for states and allow the electromagnetic decay
to be separated into isoscalar and isovector components.  For the $A$ = 10 system,
the isovector contributions are predicted to be small, symmetric, and act to
enhance the decay in $^{10}$Be and suppress it in $^{10}$C, resulting in
a $^{10}$C decay that should be $\sim$10$\%$ lower than that in $^{10}$Be. The relative $B(E$2)
values in $^{10}$C and $^{10}$Be thus represent an interesting test of nuclear modeling and of the isospin
dependence of electromagnetic decays.

Within the $A$ = 10 system, $^{10}$C is the more exotic partner. It has only one
bound excited state, with $J^{\pi}$ = $2^+$, at 3354 keV. It becomes unbound at
4006 keV at which point it can disintegrate into $^{9}$B+$p$. Its mirror
partner, $^{10}$Be, has six bound states below a breakup threshold of 6812 keV
where the $^9$Be+$n$ channel opens. The first excited state of $^{10}$Be also
has $J^{\pi}$ = $2^+$ and lies at 3368 keV, a sign that, in excitation energy at
least, these configurations are similar, despite the difference in binding
energy.

A pioneering Doppler Shift Attenuation Method (DSAM) experiment by Fisher
{\it et al.,}~\cite{fisher} in 1968 was aimed at understanding these issues and
testing the intermediate-coupling shell model predictions by measuring the decay
rates from the first excited states in $^{10}$C and $^{10}$Be. However, the DSAM
technique was new and only $\sim$20$\%$ precision could be achieved.
$^{10}$C was found to be slightly more collective, but not in glaring
disagreement with the shell model, given the large experimental
uncertainties. We have recently remeasured the lifetime of the $2_1^+$ state in
$^{10}$Be~\cite{mccutchan} and determined $B$($E$2; $2_1^+ \rightarrow 0_1^+$) =
9.2(3) $e^2$fm$^4$. Comparing this with the value $B$($E$2; $2_1^+ \rightarrow 0_1^+$) = 12.2$\pm$1.9
$e^2$fm$^4$ measured by Fisher {\it et al.,} for $^{10}$C still supports
a larger $B$($E$2) rate in $^{10}$C. However, the carbon
value has substantial experimental uncertainties, so now, with far superior
experimental tools and much refined theory, we can re-address this interesting
problem at a level of precision which should provide stringent tests of {\it ab-initio}
calculations.

\section{Experiment}

The $2_1^+$ state in $^{10}$C was populated in the inverse kinematics
$p$($^{10}$B, $n$)$^{10}$C reaction. Beams of $^{10}$B ions of $\sim$ 1pnA and
95 MeV were produced by the ATLAS accelerator at Argonne National
Laboratory. Targets consisted of thin layers of CH$_2$ on thick backings of
copper and gold.  $^{10}$C nuclei recoiling along the beam direction were
selected by the Argonne Fragment Mass Analyzer (FMA)~\cite{fma} positioned 90
cm downstream of the target and subtending $1^{\circ}$ around the beam
direction. The $^{10}$C nuclei were produced at recoil velocities of $\beta$ =
$v$/$c$ $\sim$ 13$\%$ and emerged from the backing target layer with $\beta$
$\sim$ 10$\%$.  To satisfy the FMA energy acceptance window, the recoils had to
be further slowed down to $\beta$ $\sim$ 8$\%$.  This was achieved through a
series of titanium degrader foils placed at the entrance to the FMA. $^{10}$C
ions with $q$ = 6$^+$ were transported to the focal plane while most
non-interacting beam particles were rejected by the FMA.  The selection of
$q$ = $6^+$ ions was very effective for suppressing scattered beam particles.  The
transmitted ions first passed through two PPAC detectors before being stopped 50
cm behind the focal plane in a 30-cm deep, two-electrode ionization-chamber operated
at 50 torr. Gamma rays were detected with the Gammasphere
array~\cite{gammasphere} consisting of 100 Compton suppressed HPGe detectors in
16 azimuthally symmetric rings from $\theta$ = $34^{\circ}$ to $163^{\circ}$ relative to
the beam direction.

\begin{figure}
\center{{\includegraphics[width=85mm]{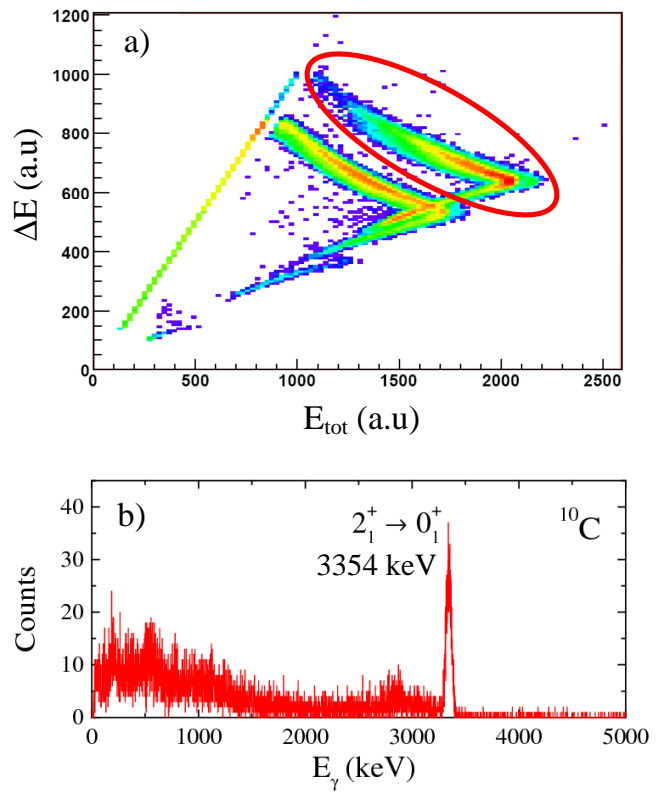}}}
\caption{(Color online) (a) Energy loss versus total energy plot of data from the
  ionization chamber behind the focal plane of the FMA.  The solid red circled
  region is direct population of the $2_1^+$ state in $^{10}$C. Data were
  obtained with a 116-$\mu$g/cm$^2$ CH$_2$ target on a 23.8-mg/cm$^2$ copper backing. (b)
  Gamma-ray spectrum obtained by gating on the excited $^{10}$C recoils.}
\label{fig-dee}
\end{figure}

Figure~\ref{fig-dee}(a) shows a typical energy loss ($\Delta E$) versus total energy
($E_{tot}$) spectrum obtained from the ionization chamber.  The locus with the
largest $\Delta E$ (solid red circled region) corresponds to direct population
of the $2_1^+$ state in $^{10}$C.  The large spread in total energy for these
recoils stems from energy scattering and straggling in the backing and degrader
foils. The wide strip of counts below the $^{10}$C recoils is identified with $^{10}$B
scattered beam.  The pressure in the ionization chamber was not sufficient to fully stop
the highest energy $^{10}$C recoils, resulting in a wrap around feature (punch
through) in the $\Delta E$ versus $E_{tot}$ plot. Direct population of the ground state
of $^{10}$C is also observed, although due to the punch through, it falls appears
in the same $\Delta E$ versus $E_{tot}$ area as the $^{10}$B
scattered beam. The $\gamma$-ray spectrum obtained by gating on the excited $^{10}$C recoils is
given in Fig.~\ref{fig-dee}(b), showing only the 3354-keV, $2_1^+ \rightarrow 0_1^+$
transition in $^{10}$C.

\begin{figure}
\center{{\includegraphics[width=85mm]{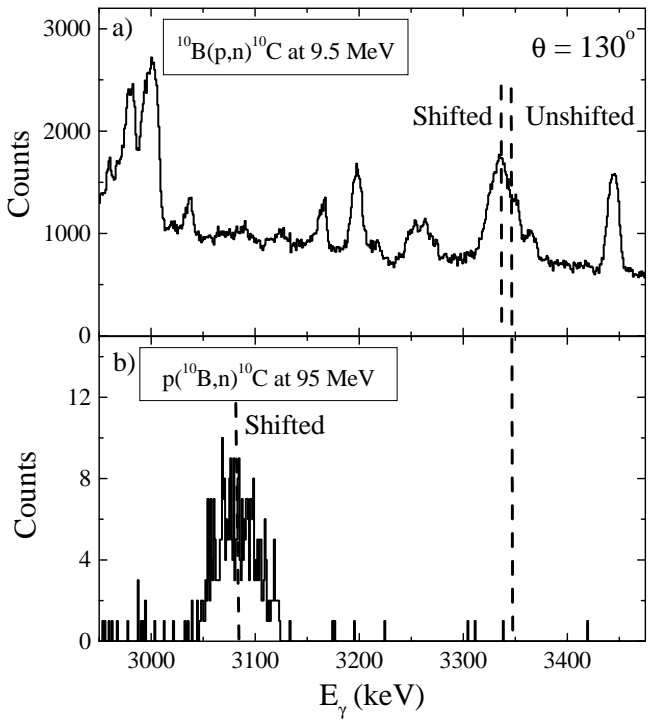}}}
\caption{Comparison of $\gamma$-ray spectra for $^{10}$C decays from DSAM in (a)
  regular and (b) inverse kinematics. Both spectra correspond to the $\theta$ =
  130$^\circ$ angle group in Gammasphere. Reaction details are included in the
  figure.}
\label{fig-gammas}
\end{figure}

The current setup offers several advantages over the prior DSAM measurement
performed with a regular-kinematics reaction. To compare the present technique
with a normal kinematics DSAM measurement, the original Fisher {\it et al.,}
experiment~\cite{fisher} was also repeated with Gammasphere.  This regular-kinematics
experiment was performed with a 9.5-MeV proton beam incident on a
$^{10}$B target followed by a gold backing.  Gamma rays were detected with
the Gammasphere array.  A comparison of the spectra obtained from the regular-
and inverse-kinematics reactions is given in Fig.~\ref{fig-gammas} for the same angle group
($\theta$ = 130$^\circ$) in Gammasphere.  The regular kinematics spectrum
(Fig.~\ref{fig-gammas}(a)) is very complicated, with considerable background from reactions of
the high-energy protons on the $^{10}$B target, the gold backing and scattering
in the target chamber.  In contrast, requiring detection of a recoiling
$^{10}$C residue by the FMA almost entirely suppresses the background
(Fig.~\ref{fig-gammas}(b)), providing a spectrum where only the $2_1^+ \rightarrow 0_1^+$
transition in $^{10}$C is observed. Selection of the $^{10}$C recoils with the
FMA also provides a well-defined angle between the recoil velocity vector and
the direction (and subsequent detection) of the $\gamma$-rays. Recoil detection comes at the cost of
poorer statistics; however, with almost no background the peak centroids can be
reliably determined to 1-2 keV.  With the regular-kinematics reaction, the
recoiling $^{10}$C ions have a kinetic energy of a few MeV, an energy regime where
stopping powers are poorly determined and the $\gamma$-ray energy shifts are
only a few keV. In the inverse-kinematics reaction, $^{10}$C nuclei were
produced at a very high recoil energy ($E$ $\sim$ 80 MeV). This allows the
measurement to be performed in a velocity regime where the stopping is 99.99$\%$
electronic and most precisely known. The high recoil velocity also produces
large Doppler shifts.  With the set of Gammasphere angles, the range of forward-shifted
to backward-shifted peaks spanned more than 800 keV.

In a DSAM measurement the lifetime is derived from the difference between
production and emission velocities.  For the inverse-kinematics reaction, the
distribution of recoil products is very forward peaked and there are two
solutions for residues moving along the initial beam direction, depending on the
direction of the emitted neutron.  For this particular experiment, recoils
emitted at $0^{o}$ relative to the beam direction can have recoil velocities of 75 or 82
MeV.  The FMA was always set to transmit the higher-energy recoil group.  The
initial $\beta$ value at production is first measured using a self-supporting CH$_2$
target and correcting for the small energy loss in the target.  Degrader foils
of the same thickness as the backed-target experiments were used to replicate the
DSAM measurements and ensure that the FMA entrance conditions were the same for
both the self-supporting and backed targets.

\begin{figure}
\center{{\includegraphics[width=85mm]{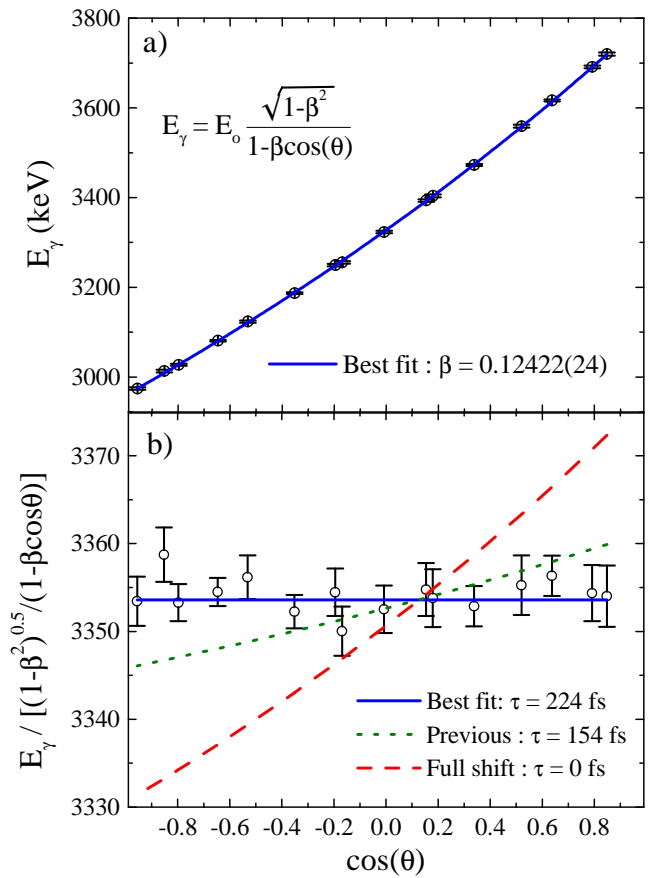}}}
\caption{(Color online) (a) Measured centroids of the 3354-keV transition for
  each Gammasphere angle group. The solid curve shows the result of the
  relativistic Doppler shift formula taking the best-fit value of $\beta$. (b)
  Similar to (a) but normalized to
  $\sqrt{1-\beta^2}$/(1-$\beta$cos($\theta$)). Lines include the best fit value
  of $\beta$ (solid line), maximum $\beta$ value allowed by the reaction
  kinematics (dashed line), and the $\beta$ value corresponding to the previous
  lifetime measurement (dotted line).}
\label{fig-angle-groups}
\end{figure}

To determine the average velocity of the recoils at the time of $\gamma$-ray emission,
the centroid of the 3354-keV, $2_1^+
\rightarrow 0_1^+$ transition was determined for each of the 16 Gammasphere angle
rings.  In Fig.~\ref{fig-angle-groups}(a), the measured centroid is plotted as a
function of cos($\theta$) for a 105-$\mu$g/cm$^2$ CH$_2$ target on a 23-mg/cm$^2$
copper backing.  A fit to these data yields an average $\beta$ at the time of $\gamma$-ray
emission of $\beta$ = 0.12422(24).  For reference, a lifetime of $\tau$ $\sim$ 0 would give $\beta$ $\sim$ 0.131
(the production velocity) while an infinitely long lifetime would yield $\beta$ $\sim$ 0.099 (velocity after
emerging from the backing layer). The measured centroids are compared to the
relativistic Doppler shift formula using the best fit $\beta$ in
Fig.~\ref{fig-angle-groups}(a).  Figure~\ref{fig-angle-groups}(b) illustrates the
quality of the fit more clearly, by dividing the measured centroids by the function
$\sqrt{1-\beta^2}$/(1-$\beta$cos($\theta$)). Included are lines for $\beta$ values for
the best fit $\beta$ ($\beta$ = 0.12422, corresponding to $\tau$ = 224 fs),
production ($\beta$ = 0.131, corresponding to $\tau$ = 0), and that
which would correspond to the previous lifetime value ($\beta$ = 0.1265 for $\tau$ =
154 fs).

To determine the lifetime of the level of interest from the measured mean
$\beta$ value, the thicknesses of the target and backing layers must be known,
as this defines the relevant transit time scale. The backing foils were prepared
by rolling gold and copper foils to the appropriate thickness. Targets were then
prepared by dipping the backing foils into solutions of xylene and C$_2$H$_2$,
and subsequently evaporating the xylene.  The thin layer of CH$_2$ on a very
thick backing of gold or copper made a precise measurement of the CH$_2$
thickness by a traditional $\alpha$-gauging method impossible. The CH$_2$
thickness was determined by comparing the yield (number of $^{10}$C recoils
detected in the ionization chamber, per beam intensity, per time) of a backed
target relative to a commercially made self-supporting CH$_2$ target of known
thickness.  A series of degrader foils, identical to the target backing, were
placed behind the self-supporting CH$_2$ target to achieve the same energy
recoils into the FMA as with the backed targets. The thickness of the CH$_2$
layer on the backed targets was measured before and after the DSAM measurement
via a relative yield measurement as described above.  An approximately 20$\%$
reduction in yield was observed.  For the DSAM analysis, the average of the
thickness before and after the measurement was used and the full range of
thickness taken as the uncertainty.  As the target layers are thin, do not
provide much stopping and the ions move very quickly through them, the target
thickness does not contribute greatly to the systematic uncertainty.  The
characterization of carbon and boron ions slowing in CH$_2$, copper and gold was
taken from the SRIM~\cite{srim} and MSTAR~\cite{mstar} packages. The two models
differ in stopping powers on the order of 3$\%$ in the relevant velocity regime.
These differences were incorporated into the systematic uncertainty.

\begin{table}
\caption{Mean lifetimes from different target and backing combinations
  determined for the 3354-keV level in $^{10}$C. The CH$_2$ thicknesses are
  the average of the measured values at the beginning and end of each DSAM measurement.}
\begin{tabular}{ccccccc}
\hline Target (CH$_2$) & $\;$ & Backing & $\;$ & $\tau$ & & $\;$ $\Delta\tau_{stat}$ \\
($\mu$g/cm$^2$) & & (mg/cm$^2$) & & (fs)& $\;$ & (fs) \\
\hline
105 & & 23.0 Cu & & 224 & & $\pm$8 \\
150 & & 31.0 Au & & 215 & & $\pm$10 \\
170 & & 23.8 Cu & & 219 & & $\pm$13 \\
80  & & 24.0 Au & & 198 & & $\pm$18 \\
300 & & 14.9 Au & & 216 & & $\pm$23 \\
\hline
\end{tabular}
\end{table}

The lifetime of the 3354-keV level was measured in five separate experiments.
The target characteristics and the extracted lifetimes are summarized in Table
I. The weighted mean value is $\tau$= 219${\pm}$(5)$_{stat}$
${\pm}$(10)$_{syst}$fs which implies $B$($E$2; $2_1^+ \rightarrow 0_1^+$) =
8.8(3) $e^2$fm$^4$. This lifetime is substantially longer than the previous value
obtained by Fisher {\it et al.,}~\cite{fisher}.  Figure~\ref{fig-be2-exp} gives a comparison of
the current measurements (solid symbols) and those of Ref.~\cite{fisher} (open
symbols). Clearly, the data are now much better constrained for investigating
the symmetry of the wave functions, but theoretical guidance is needed to infer
the meaning of the result.

\section{Theory}

\begin{figure}[b]
\center{{\includegraphics[width=85mm]{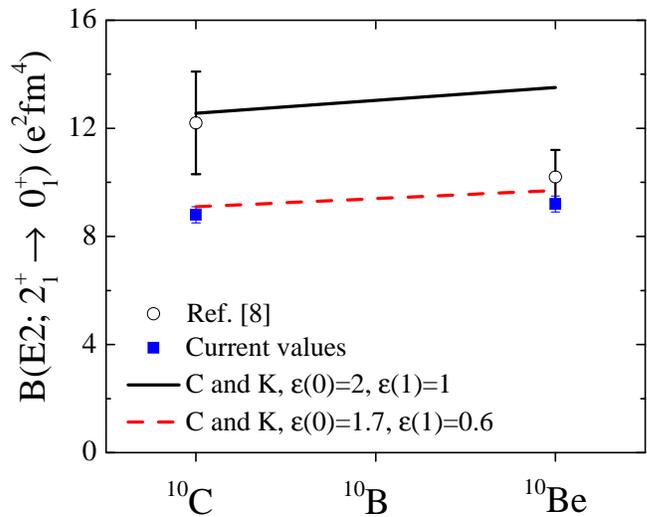}}}
\caption{(Color online) $B$($E$2; $2_1^+ \rightarrow 0_1^+$) transition
  strengths (in $e^2$fm$^4$) for $^{10}$C and $^{10}$Be. Open symbols are the
  results of Ref.~\cite{fisher} while solid symbols are the current work and the
  recent measurement of Ref.~\cite{mccutchan}.}
\label{fig-be2-exp}
\end{figure}

Empirically, assuming charge symmetry for the wave functions, the transition
strengths can be written as
\begin{equation}
B(E2) = [M(E2)]^2/5 = [AT + BT_z]^2
\end{equation}
where $M(E2)$ is the reduced matrix element and we use the convention that $T_z =
+\frac{1}{2}$ for the neutron.  The new $^{10}$Be and $^{10}$C data can be used to
infer that the isoscalar term is dominant, $A$ = 3.00(1) $e$fm$^2$, while the
isovector term is much smaller, $B$ = 0.03(3) $e$fm$^2$, a 1$\%$ effect.  In
conventional shell-model calculations, isospin enhancements or effective charges,
$\epsilon(T)$ are introduced to account for effects such as core polarization:
$A=A^\prime\epsilon(0)$ and  $B=B^\prime\epsilon(1)$, where $A^\prime$ and $B^\prime$ are
constants derived for a particular wave function.  Very early shell model
calculations of Cohen and Kurath~\cite{cohenandkurath,alburger} for the $A$ = 10
system gave predictions for $B$($E$2) strengths in the form of Eq. (1).  These are
included in Fig.~\ref{fig-be2-exp} (solid black line).  These $p$-shell, mirror-
symmetric wave functions provide the correct slope for describing the transition
strengths between $^{10}$C and $^{10}$Be, however, overestimate the overall magnitude
(Fig.~\ref{fig-be2-exp}) due to the use of very simple isoscalar and isovector
enhancements, $\epsilon$(0) = 2 and $\epsilon$(1) = 1. Using isoscalar and isovector
enhancements now broadly accepted for $p$-shell calculations~\cite{eff1,eff2},
$\epsilon$(0) = 1.7 and $\epsilon$(1) = 0.6, one obtains $^{10}$Be $B$($E$2; $2_1^+ \rightarrow 0_1^+$) = 9.7
$e^2$fm$^4$ and $^{10}$C $B$($E$2; $2_1^+ \rightarrow 0_1^+$) = 9.1 $e^2$fm$^4$, close to
the experimental results. These calculations are included in Fig.~\ref{fig-be2-exp} (dotted red line).

There are two ways to interpret the dominant isoscalar term along with a nearly
zero isovector contribution to the $B$($E$2) strengths.  Conceptually, this may
be perceived as the two-alpha cluster in these nuclei being the same in
$^{10}$Be and $^{10}$C and their respective valence particles contributing
little.  This mirror-symmetric interpretation conforms well to the predictions
of the standard shell model.  One can alternately consider the possibility that mirror
symmetry is not preserved; for example, the alpha-cluster spacing is modified, but this
effect is offset by differing contributions from the valence particles. This
effect can be investigated through the use of more sophisticated models.

The variational Monte Carlo (VMC) and Green's function Monte Carlo (GFMC)
methods have been very useful in improving our understanding of light
nuclei~\cite{bobandsteve1,bobandsteve2}, and successful in reproducing the
electric quadrupole collectivity in $^{10}$Be, without resorting to the use of any effective
charges.  Using realistic two- and three-body forces and operators
(including explicit charge-symmetry-breaking terms), this is a good approach for
exploring $^{10}$C and the symmetry of the $A$ = 10 wave functions.  More details
on the GFMC method of calculating transition strengths are given in
Ref.~\cite{pervin}.

\begin{figure}
\center{{\includegraphics[height=85mm,angle=270]{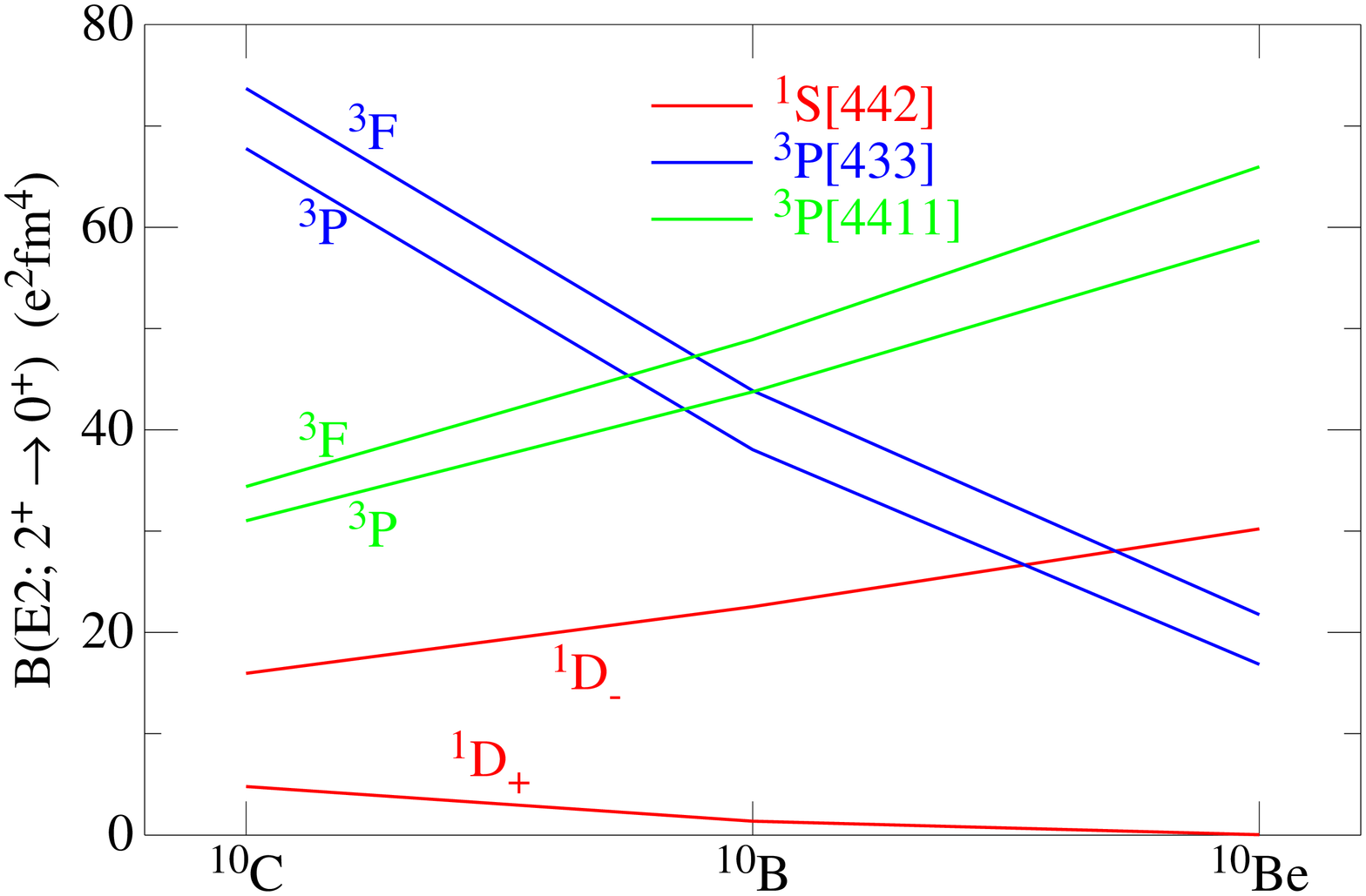}}}
\caption{(Color online) $B(E$2) transition strengths (in $e^2$fm$^4$) for
VMC basis state pairs. Only transitions that conserve spatial
symmetry are shown.  The legend gives the designation of the 0$^+$ states
while the curves are labeled with the $^{2S\!+\!1}L$ of the 2$^+$ states.}
\label{fig-individual}
\end{figure}

The VMC calculations use trial wave functions containing
non-central, two- and three-body correlation operators acting on an
antisymmetrized one-body wave function, $\Phi(JMTT_z)$, which determines
the quantum numbers of the state being computed.  The $\Phi(JMTT_z)$ wavefunction is
expanded in $LS$-basis functions~\cite{bobandsteve2}:

\begin{widetext}

\begin{equation}
\Phi(JMTT_z) = \sum_{LS[n]} \beta(^{2S\!+\!1}L[n],JTT_z)\Phi(^{2S\!+\!1}L[n],JMTT_z) ,
\end{equation}
where the amplitudes $\beta(^{2S\!+\!1}L[n],JTT_z)$ are found from a diagonalization
of the Hamiltonian.  For $^{10}$Be, we construct states from the three highest
spatial symmetries as denoted by the Young diagram $[n]$ (see Ref.~\cite{W06}).
This gives three basis functions for the 0$^+$ ground state:
$^1$S[442], $^3$P[4411], and $^3$P[433], while the
2$^+$ states have six basis functions: $^1$D[442]$_-$, $^1$D[442]$_+$,
$^3$P[4411], $^3$P[433], $^3$F[4411], and $^3$F[433].  Note that there
are two linearly independent $^1$D[442] basis states; the distinction between
them is arbitrary and we choose to express them as eigenfunctions of
the quadrupole operator, with the subscript indicating the sign of the
quadrupole moment.  The VMC $E2$ matrix element is
\begin{eqnarray}
M(E2) = \sum_{L^{\prime}S^{\prime}[n^{\prime}],LS[n]} &&
\beta(^{2S^{\prime}\!+\!1}L^{\prime}[n^{\prime}],J\!=\!0,T\!=\!1,T_z) \beta(^{2S\!+\!1}L[n],J\!=\!2,T\!=\!1,T_z)  \nonumber \\
&& \times \langle {\cal C} \Phi(^{2S^{\prime}\!+\!1}L^{\prime}[n^{\prime}],J\!=\!0,T\!=\!1,T_z) || E2
|| {\cal C} \Phi(^{2S\!+\!1}L[n],J\!=\!2,T\!=\!1,T_z) \rangle \ ,
\label{eq:me2}
\end{eqnarray}
\end{widetext}
where ${\cal C}$ denotes the two- and three-body correlations and there are
3$\times$6 contributions to the sum.  The $E2$ operator does not change
spatial symmetry, so the only big contributions are those from $^1$D[442]$_-$ or $^1$D[442]$_+$
to $^1$S[442], $^3$P[4411] or $^3$F[4411] to $^3$P[4411], and $^3$P[433] or $^3$F[433] to $^3$P[433]
(the ${\cal C}$ do not conserve the spatial symmetry so there are small non-zero
matrix elements for the other possibilities).  These individual contributions,
calculated with wave functions for the AV18 two-nucleon
and Illinois-7 three-nucleon potentials (AV18+IL7)~\cite{WSS95,P08},
are shown in Fig.~\ref{fig-individual} for isospin-symmetric basis states, i.e.,
the parameters in $\Phi(^{2S\!+\!1}L[n],J\!=\!2,M,T\!=\!1,T_z)$ are independent of $T_z$.
The diagonalization of the two $^1$D[442] states into the quadrupole basis
was done for $^{10}$Be and not changed for $^{10}$B and $^{10}$C.
As can be seen, these isospin-symmetric calculations can give very different
$T_z$ behaviors, depending on the pair of $^{2S\!+\!1}L[n]$ states being used.
Calculations with basis states containing different variational parameters
give very similar results to those in the figure; we believe that the trends
shown result from the different $^{2S\!+\!1}L[n]$ values of the pairs and thus also would
be obtained with other realistic Hamiltonians and even for the corresponding
harmonic-oscillator shell model states.
This means that the nearly $T_z$-independent $B(E2)$ strengths obtained in the shell
model calculation require a specific combination of states.

\begin{figure}
\center{{\includegraphics[height=85mm,angle=270]{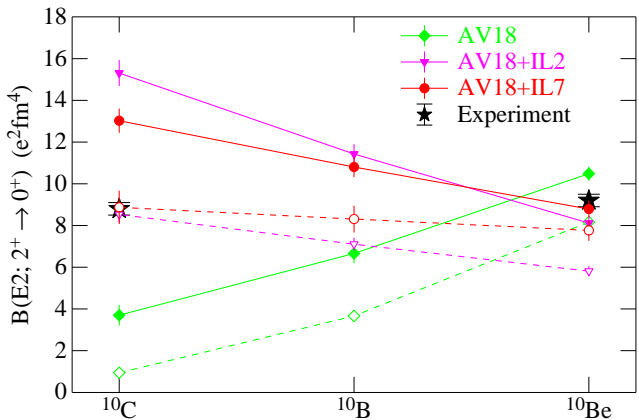}}}
\caption{(Color online) VMC and GFMC calculations of $B(E2)$ strengths for the 2$^+_1$
($Q<0$) state assuming isospin symmetric wave functions. The wave functions
were computed for $^{10}$Be using the indicated Hamiltonians.  VMC results are shown
as open symbols and dashed lines; GFMC results are solid symbols and solid lines.}
\label{fig-isosym_vmc_gfmc}
\end{figure}

Figure~\ref{fig-isosym_vmc_gfmc} shows VMC and GFMC calculations of $B(E2)$ values for the 2$^+_1$
($Q<0$) state assuming isospin symmetric wave functions. The wave functions
were computed for $^{10}$Be using the AV18 interaction alone or with
the IL2 or IL7 three-body potentials.
As shown in Ref.~\cite{mccutchan}, the  2$^+_1$ state of $^{10}$Be has a negative
quadrupole moment and a strong $E2$ decay to the ground state for
all the Hamiltonians.  (For AV18 alone, the energies of the two $2^+$ states are nearly
degenerate, so we choose to identify the $Q<0$ state as the 2$^+_1$ state.)
The reduced matrix elements for $^{10}$C were obtained by interchanging protons
and neutrons in the $^{10}$Be wave functions and the $^{10}$B reduced
matrix elements are the average of the $^{10}$Be and $^{10}$C, i.e.,
the wave functions are isospin symmetric with those of $^{10}$Be.  There
is considerable variation in the $T_z$ behavior of the $B(E2)$ strengths for the different
Hamiltonians in the VMC calculations.  This is presumably due to the different
$\beta(^{2S\!+\!1}L[n],J\!=\!2,T\!=\!1,T_z=+1)$ amplitudes from the separate diagonalizations.
The GFMC generally preserves, or even enhances, these different trends which
suggests a strong sensitivity of the isovector $B(E2)$ to the three-body force.

\begin{figure}
\center{{\includegraphics[height=85mm,angle=270]{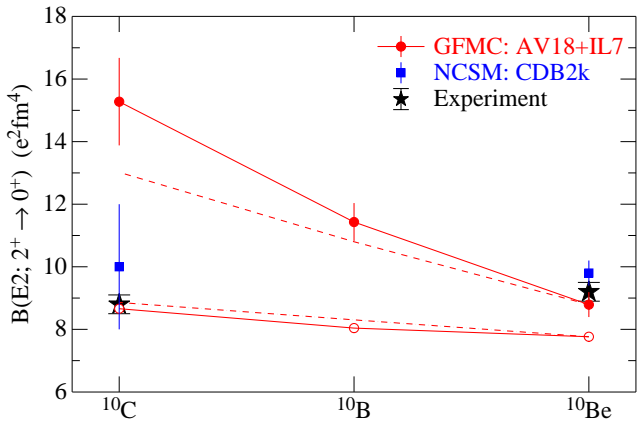}}}
\caption{(Color online) VMC and GFMC calculations of $B(E2)$ for the lowest
($Q<0$) 2$^+$ state with the AV18+IL7 Hamiltonian.  The wave functions
were computed separately for each nucleus using different VMC correlations
and independent GFMC propagation.  VMC results are shown
as open symbols; GFMC results are solid symbols. The dashed lines show the corresponding
isospin symmetric results from Fig.~\ref{fig-isosym_vmc_gfmc}.}
\label{fig-independent_vmc_gfmc}
\end{figure}

Isospin symmetry of the wave functions is certainly only an approximation.
Due primarily to the increasing Coulomb potential energy going from $^{10}$Be
to $^{10}$C, the $^{10}$C states under consideration are less bound (0.5 MeV
vs. 3.5 MeV for the $2_1^+$ state) and,
hence, should be more diffuse.  This can be studied by performing separate
calculations for each nucleus.  We have done such calculations for
the AV18+IL7 Hamiltonian.  The one-body parts of the VMC wave functions
are solutions of Woods-Saxon wells plus an average Coulomb potential~\cite{bobandsteve2};
the strength of the Coulomb term is proportional to the number of $p$-shell
protons.  Separate diagonalizations were made for each nucleus, so the
$\beta$ are also different.  The GFMC propagations are still made
in a good isospin basis, but the isoscalar Coulomb potential used
reflects the total charge of the nucleus~\cite{pervin}.  The results
of these calculations are compared with the isospin-symmetric AV18+IL7
calculations in Fig.~\ref{fig-independent_vmc_gfmc}.  The independent
calculations for $^{10}$C and $^{10}$B are not very different from
the isospin-symmetric extrapolations from the $^{10}$Be results.
Unfortunately, the already too large value for the $^{10}$C B$(E2)$ is
further increased.

In light of the apparent failure of these GFMC calculations to reproduce the
$B$($E$2; $2_1^+ \rightarrow 0_1^+$) transition strength in $^{10}$C, it is
important to consider possible shortcomings of the calculation which could
cause the discrepancy.
One possibility is that, with the weaker binding of the $2_1^+$ state in
$^{10}$C compared to $^{10}$Be, contributions from
beyond the $p$-shell might become important.
In fact, the VMC trial functions already have a fair admixture
of $sd$-shell and higher components due to two-body tensor correlations
in the $\cal C$ of Eq.~\ref{eq:me2} (see Ref.~\cite{bobandsteve1}),
and these are further enhanced in the GFMC propagation.
However, to further test this possibility, we constructed alternative
clusterized VMC trial functions with explicit $sd$-shell components.
These wave functions combine a $^8$Be($0^+$ or $2^+$) core with two final
nucleons in $p$- or $sd$-shell orbitals (with appropriate Coulomb terms),
all $LS$-coupled to give the appropriate total $J^\pi$.
The $0^+$ ($2^+$) states have four (seven) $p$-shell and eight (seven)
$sd$-shell components; separate diagonalizations for the corresponding
$\beta$s are made for each $T_z$.
For a Hamiltonian containing AV18 and the Urbana IX (UIX) three-body
potential~\cite{PPCW95}, the $sd$-shell $\beta$s contribute only
2.5\% of the total wave function in the $^{10}$Be $0_1^+$ state, but
21\% in the $0_2^+$ state; these numbers increase to 3.4\% and 28.9\%,
respectively, in $^{10}$C.
The $2_1^+$ and $2_2^+$ states both have 4\% or less $sd$-shell contributions,
with only slightly greater amounts in $^{10}$C than in $^{10}$Be.
(Interestingly, 96\% of the $2_1^+$ state has a $^8$Be($2^+$) core in
this construction, clearly indicating it is the $J$=2 member of the $K$=0
rotational band.)
Consequently, inclusion of the $sd$-shell components has only a very minor
effect on the $B(E2)$ values; in $^{10}$Be using $p$-shell only components
gives $B(E2)$ = 8.6 $e^2$fm$^4$, while adding $sd$-shell components
raises it to 9.2 $e^2$fm$^4$.
The corresponding $B(E2)$ values in $^{10}$C are 9.6 and 11.6 $e^2$fm$^4$,
respectively, showing the same moderate change with $T_z$ as the AV18+IL2 and
AV18+IL7 Hamiltonians.

\begin{table*}
\caption{GFMC calculations of the $A$ = 10, $T$ = 1, ground-state energies
$E_{gs}$ and excitation energies $E_x$ in MeV for several
Hamiltonians used in this work, plus NCSM results for the CD-Bonn 2000
potential and experimental values. ($^{10}$B excitations are shown
relative to the $0^+;1$ isobaric analog.)
Also shown are ground state charge radii in fm, quadrupole moments
for the excited states in $e{\rm fm}^2$, and the $B(E2)$ transition
strengths in $e^2{\rm fm}^4$.  Asterisks denote GFMC results obtained
with isospin-symmetric wave functions generated from $^{10}$Be (see text).}
\begin{ruledtabular}
\begin{tabular}{ccddddd}
& & \multicolumn{1} {c}{NCSM} &\multicolumn{3} {c}{GFMC} & \multicolumn{1} {c}{Expt.} \\
\multicolumn{1} {c}{$^AZ$} &
\multicolumn{1} {c}{Observable} &
\multicolumn{1} {c}{CDB2k} &
\multicolumn{1} {c}{AV18} &
\multicolumn{1} {c}{AV18+IL2} &
\multicolumn{1} {c}{AV18+IL7} & \\
\hline
$^{10}$Be  & $|E_{gs}(0^+)|$& 56.5(5) & 50.1(1) & 66.4(4) & 64.1(3) & 64.98 \\
           & $E_x(2^+_1)$   &  3.6(1) &  2.9(1) &  5.0(4) &  3.4(3) &  3.37 \\
           & $E_x(2^+_2)$   &  4.8(1) &  3.0(1) &  5.8(4) &  5.3(3) &  5.96 \\
           & $r_c$          &  2.25(5)&  2.47(1)&  2.33(1)&  2.33(1)&  2.36(2) \\
           & $Q(2^+_1)$     & -5.9(5) & -4.1(1) & -4.9(1) & -6.7(1) &  \\
           & $Q(2^+_2)$     &  5.3(5) &  5.8(1) &  0.2(1) &  4.5(1) &  \\
& $B(E2;2^+_1\rightarrow0^+)$
                            &  9.8(4) & 10.5(4) &  8.1(3) &  8.8(4) &  9.2(3) \\
& $B(E2;2^+_2\rightarrow0^+)$
                            &  0.2(2) &  3.4(2) &  3.3(2) &  1.8(1) &  0.11(2) \\
\hline
$^{10}$B   & $|E(0^+;1)|$   & 55.3(5) & 48.3(3)*& 64.6(4)*& 62.6(2) & 63.01 \\
           & $E_x(2^+_1;1)$ &         &  2.9(4)*&  5.0(5)*&  3.6(3) &  3.42 \\
           & $E_x(2^+_2;1)$ &         &  3.0(4)*&  5.8(5)*&  5.2(5) &       \\
           & $Q(2^+_1;1)$   &         & -5.8(1)*& -3.5(1)*& -2.7(1) &  \\
           & $Q(2^+_2;1)$   &         &  7.9(1)*& -2.0(1)*&         &  \\
& $B(E2;2^+_1;1\rightarrow0^+;1)$
                            &         &  6.7(5)*& 11.4(5)*& 11.4(6) &         \\
& $B(E2;2^+_2;1\rightarrow0^+;1)$
                            &         &  8.9(4)*&  1.0(1)*&         &         \\
\hline
$^{10}$C   & $|E_{gs}(0^+)|$& 51.9(5) & 45.8(3)*& 61.7(4)*& 60.0(2) & 60.32 \\
           & $E_x(2^+_1)$   &  3.6(1) &  2.7(3)*&  4.7(4)*&  3.2(3) &  3.35 \\
           & $E_x(2^+_2)$   &  4.3    &  2.8(3)*&  5.4(4)*&  5.1(5) &       \\
           & $r_c$          &         &  2.77(1)*& 2.55(1)*& 2.65(1)&       \\
           & $Q(2^+_1)$     & -1.1(12)&-7.5(2)*& -2.1(2)*& -2.7(2) &  \\
           & $Q(2^+_2)$     &         & 10.0(2)*& -4.2(2)*& -0.9(3) &  \\
& $B(E2;2^+_1\rightarrow0^+)$
                            & 10(2) &  3.7(5)*& 15.3(6)*& 15.3(1.4)& 8.8(3) \\
& $B(E2;2^+_2\rightarrow0^+)$
                            &         & 17.0(8)*&  0.0(1)*&  0.2(1) &         \\
\end{tabular}
\end{ruledtabular}
\label{tab:qmc}
\end{table*}

A more likely possibility is simply that the Hamiltonians tested are not
adequate for these transitions.
The first priority in theoretically modeling the nuclear Hamiltonian
has been obtaining good energies for the states in question -- both
absolute binding energy of the nucleus and excitation energies of the
higher states.
These energies are shown in Table~\ref{tab:qmc} for the various Hamiltonians
used in the present GFMC calculations, along with charge radii, quadrupole
moments, and the $B(E2)$ values.
The AV18+IL7 Hamiltonian gives a particularly good overall reproduction
of both absolute binding and excitation energies.
The charge radius for $^{10}$Be is also in excellent agreement
with a recent measurement~\cite{Be10-rms}.
Table~\ref{tab:qmc} and Fig.~\ref{fig-isosym_vmc_gfmc} indicate that while
the various models tested give rather similar results for the $B(E2)$ strengths
in $^{10}$Be, they give a much more widely varying range of results for $^{10}$C.
The quadrupole moments also have far more variation in $^{10}$C.

In Table~\ref{tab:qmc} we also show results for the CD-Bonn 2000 (CDB2k)
potential~\cite{CDB2k} evaluated with the no-core shell model (NCSM)
\cite{NCSM-10Z,NCSM-10C,Nav11}.
These $B(E2)$ results are in excellent agreement for both the transitions
in $^{10}$Be and also very good for $^{10}$C, although they note that
separating the two $2^+$ states in $^{10}$C is non-trivial, leading to
large error bars for the transition and quadrupole moment.
The ground state energies are significantly underbound, which is not
surprising for a two-nucleon force alone, but the $2^+$ states
are reasonably well separated.  The $^{10}$Be charge radius is also
too small, probably due to the practical limitations of
using a finite harmonic-oscillator-space basis.

Many other groups are developing new methods for calculating light nuclei. They range
from models based on effective field theories~\cite{epelbaum} and density functionals~\cite{yao}, to monte
carlo shell models~\cite{liu} and cluster-based approaches~\cite{enyo,neff}. New, precise experimental measurements
of masses, radii, moments and decays can all help to refine these approaches and improve our insight
into nuclear structure and its evolution with neutron-to-proton ratio.

\section{Conclusions}

We have performed a precise measurement of the lifetime of the first (and
only) excited bound state in $^{10}$C. The new measurement implies a matrix
element that is considerably smaller than previously reported. It is only 2$\%$
different from its mirror transition in $^{10}$Be.
GFMC calculations with our best Hamiltonian fail to reproduce this near
equality.  Different basis states have very different trends of the
$B(E2)$ going across this isomultiplet.  Thus, if the wave functions
are nearly isospin symmetric, our calculations fail to get
the correct mixture of states.  Inclusion of explicit $sd$-shell amplitudes
into the trial wave functions, while slightly different in $^{10}$Be and $^{10}$C,
appears to be a small effect that does not resolve the discrepancy.
The calculations show only a small
amount of isospin symmetry breaking.  The definite experimental signal of
such symmetry breaking for this isovector transition operator would be an
observation that the matrix element for the transition between the
first $T=1$, 2$^+$ and 0$^+$ states in $^{10}$B is not the average of
those for $^{10}$Be and $^{10}$C. Such a measurement is underway.

Obtaining a better theoretical result would appear to require a Hamiltonian
that produces a very specific combination of spatial symmetry components.
It may be possible to develop improved three-nucleon potentials
that provide both good energies and transition strengths at the same time.
Recently, it has been suggested that using the weak $\beta$-decay of tritium
in addition to the three-nucleon binding energies is a good way to select
an optimal combination of low-energy constants in chiral N3LO three-nucleon
potentials~\cite{Gazit09}.
In a similar manner, electromagnetic transitions like the ones studied
here may provide a valuable constraint on suitable Hamiltonians, i.e. on the 3-body term.

\acknowledgments

Understanding our measurement for $^{10}$C has been challenging and we have had
lengthy discussions with many people to try to understand its significance.  We
thank John Millener for discussions about the traditional $p$-shell model and
Petr Navr\'{a}til for NCSM predictions.
Calculations were performed on the parallel computers of the Laboratory
Computing Resource Center and of the Mathematics and Computer Science
Division, Argonne National Laboratory.  This work was supported by the
DOE Office of Nuclear Physics under contracts DE-AC02-06CH11357 and
DE-AC02-98CH10946, grant DE-FG02-94-ER40834 and SciDAC grant DE-FC02-07ER41457.


\begin {thebibliography}{99}

\bibitem{wuosmaa} A. H. Wuosmaa {\it et al.,} Phys. Rev. Lett {\bf 94}, 082502 (2005).

\bibitem{mueller} P. Mueller {\it et al.,} Phys. Rev. Lett. {\bf 99}, 252501 (2007).

\bibitem{grinyer} G. F. Grinyer {\it et al.,} Phys. Rev. Lett. {\bf 106}, 162502 (2011).

\bibitem{mccutchan} E. A. McCutchan {\it et al.,} Phys. Rev. Lett. {\bf 103}, 192501 (2009).

\bibitem{BM} A. Bohr and B. R. Mottleson, {\it Nuclear Structure} (World Scientific, Singapore, 1998).

\bibitem{cohenandkurath} S. Cohen and D. Kurath, Nucl. Phys. {\bf A} {\bf 101}, 1 (1967).

\bibitem{millener} D. J. Millener, Nucl. Phys. {\bf A 693}, 394 (2001) and private communication (2009).

\bibitem{fisher} T. R. Fisher, S. S. Hanna, D. C. Healey, and P. Paul, Phys. Rev. {\bf 176}, 1130 (1968).

\bibitem{fma} C. N. Davids {\it et al.,} Nucl. Instr. Meth. Phys. Res. B {\bf 70}, 358 (1992).

\bibitem{gammasphere} I-Yang Lee, Nucl. Phys. {\bf A 520}, 641c (1990).

\bibitem{srim} J. F. Ziegler, J. P. Biersack, and M. D. Ziegler, SRIM:The Stopping
  of Ions in Matter, Lulu Press, Morrisville, North Carolina (2008)
  http:$\backslash \backslash$www.srim.org

\bibitem{mstar} H. Paul and A. Schinner, Nucl. Instr. Meth. Phys. Res. B {\bf 195}, 166 (2002).

\bibitem{alburger} D. E. Alburger, E. K. Warburton, A. Gallmann, and D. H. Wilkinson, Phys. Rev. {\bf 185}, 1242 (1969).

\bibitem{eff1} B. A. Brown and B. H. Wildenthal, Ann. Rev. Nucl. Part. Sci. {\bf 38}, 29 (1988).

\bibitem{eff2} A. Umeya, G. Kaneko, and K. Muto, Nucl. Phys. {\bf A 829}, 13 (2009).

\bibitem{bobandsteve1} Steven C. Pieper and R. B. Wiringa, Ann. Rev. Nucl. Part. Sci. {\bf 51}, 53 (2001).

\bibitem{bobandsteve2} Steven C. Pieper, K. Varga, and R. B. Wiringa, Phys. Rev. C {\bf 66}, 044310 (2002).

\bibitem{pervin} M. Pervin, Steven C. Pieper, and R. B. Wiringa, Phys. Rev. C {\bf 76}, 064319 (2007).

\bibitem{W06} R. B. Wiringa, Phys. Rev. C {\bf 73}, 034317 (2006).

\bibitem{WSS95} R. B. Wiringa, V. G. J. Stoks, and R. Schiavilla, Phys. Rev. C {\bf 51}, 38 (1995).

\bibitem{P08} Steven C. Pieper, AIP Conf. Proc. {\bf 1011}, 143 (2008).

\bibitem{PPCW95} B. S. Pudliner, V. R. Pandharipande, J. Carlson, and R. B. Wiringa, Phys. Rev. Lett. {\bf 74}, 4396 (1995).

\bibitem{Be10-rms} W. N\"{o}rtersh\"{a}user {\it et al.}, Phys. Rev. Lett. {\bf 102}, 062503 (2009).

\bibitem{CDB2k} R. Machleidt, Phys. Rev. C {\bf 63}, 024001 (2001).

\bibitem{NCSM-10Z} E. Caurier, P. Navr\'{a}til, W. E. Ormand, and J. P. Vary, Phys. Rev. C {\bf 66}, 024314 (2002).

\bibitem{NCSM-10C} C. Forss\'{a}en,  R. Roth, and P. Navr\'{a}til, arXiv:1110.0634v2 (2011).

\bibitem{Nav11} P. Navr\'{a}til, private communication (2011).

\bibitem{Gazit09} D. Gazit, S. Quaglioni, and P. Navr\'{a}til, Phys. Rev. Lett. {\bf 103}, 102502 (2009).

\bibitem{epelbaum} E. Epelbaum, H Krebs, D. Lee, and  U-G Meissner,  Phys. Rev. Lett. {\bf 106}, 192501 (2011).

\bibitem{yao} J.M. Yao, J. Meng, P. Ring, Z.X. Li, Z.P. Li, and K. Hagino, Phys. Rev. C {\bf 84}, 024306 (2011).

\bibitem{liu} L. Liu, T. Otsuka, N. Shimisu, Y. Utsuno, and R. Roth, http://arxiv.org/abs/1105.2983 and to be published in PRL (2012).

\bibitem{enyo} Y. Kanada En'yo, Phys. Rev. C {\bf 84}, 024317 (2011).

\bibitem{neff} T. Neff, H. Feldmeier, and R. Roth, Nucl. Phys. {\bf A 752}, 321c (2005).

\end {thebibliography}
\end {document}